\def\~{{$\tilde{\phantom{a}}$}}

\documentclass [12pt] {article}
\usepackage{epsfig}
\usepackage{color}

\def\thebibliography#1{\section{References}\markboth
 {REFERENCES}{REFERENCES}\list
 {[\arabic{enumi}]}{\settowidth\labelwidth{[#1]}\leftmargin\labelwidth
 \advance\leftmargin\labelsep
 \usecounter{enumi}}
 \def\newblock{\hskip .11em plus .33em minus -.07em}
 \sloppy
 \sfcode`\.=1000\relax}
\def\upcite#1{\raise6pt\hbox{\scriptsize
\cite{#1}}}
  \def\lsim{\mathrel {\vcenter {\baselineskip 0pt \kern 0pt
    \hbox{$<$} \kern 0pt \hbox{$\sim$} }}}
    \def\gsim{\mathrel {\vcenter {\baselineskip 0pt \kern 0pt
    \hbox{$>$} \kern 0pt \hbox{$\sim$} }}}

\def\hline{\noalign{\hrule \vskip2pt}}

\def\|{\ifmmode\Vert\else \char`\|\fi}
\ifx\oldzeta\undefined                          
  \let\oldzeta=\zeta                            
  \def\zzeta{{\raise 2pt\hbox{$\oldzeta$}}}     
  \let\zeta=\zzeta                              
\fi

\ifx\oldchi\undefined                           
  \let\oldchi=\chi                              
  \def\cchi{{\raise 2pt\hbox{$\oldchi$}}}       
  \let\chi=\cchi                                
\fi

\def\frac#1#2{{#1 \over #2}}

\def\half{\ifinner {\scriptstyle {1 \over 2}}
   \else {1 \over 2} \fi}

\def\simge{\mathrel{%
   \rlap{\raise 0.511ex \hbox{$>$}}{\lower 0.511ex \hbox{$\sim$}}}}
\def\simle{\mathrel{
   \rlap{\raise 0.511ex \hbox{$<$}}{\lower 0.511ex \hbox{$\sim$}}}}

\def\buildchar#1#2#3{{\null\!                   
   \mathop#1\limits^{#2}_{#3}                   
   \!\null}}                                    
\def\overcirc#1{\buildchar{#1}{\circ}{}}

\def\slashchar#1{\setbox0=\hbox{$#1$}           
   \dimen0=\wd0                                 
   \setbox1=\hbox{/} \dimen1=\wd1               
   \ifdim\dimen0>\dimen1                        
      \rlap{\hbox to \dimen0{\hfil/\hfil}}      
      #1                                        
   \else                                        
      \rlap{\hbox to \dimen1{\hfil$#1$\hfil}}   
      /                                         
   \fi}                                         %

\def\subrightarrow#1{
  \setbox0=\hbox{
    $\displaystyle\mathop{}
    \limits_{#1}$}
  \dimen0=\wd0
  \advance \dimen0 by .5em
  \mathrel{
    \mathop{\hbox to \dimen0{\rightarrowfill}}
       \limits_{#1}}}                           

\def\overlay#1#2{\ifmmode%
\setbox0=\hbox{$#1$}%
\setbox1=\hbox to\wd0{\hss$#2$\hss}\else%
\setbox0=\hbox{#1}%
\setbox1=\hbox to\wd0{\hss#2\hss}\fi%
#1\hskip-\wd0\box1 }

\def\pmb#1{\leavevmode\setbox0=\hbox{#1}%
\kern-.02em\copy0\kern-\wd0
\kern.04em\copy0\kern-\wd0
\kern-.02em\raise.04em\box0 }

\def\vereq#1#2{\lower3pt\vbox{\baselineskip1.5pt \lineskip1.5pt
\ialign{$\m@th#1\hfill##\hfil$\crcr#2\crcr\sim\crcr}}}

\def\tensor#1{\protect\@ontopof{#1}{\leftrightarrow}{1.15}\mathord{\box2}}
\def\overstar#1{\protect\@ontopof{#1}{\ast}{1.15}\mathord{\box2}}
\def\overdots#1{\protect\@ontopof{#1}{\cdots}{1.0}\mathord{\box2}}
\def\overcirc#1{\protect\@ontopof{#1}{\circ}{1.2}\mathord{\box2}}
\def\loarrow#1{\protect\@ontopof{#1}{\leftarrow}{1.15}\mathord{\box2}}
\def\roarrow#1{\protect\@ontopof{#1}{\rightarrow}{1.15}\mathord{\box2}}

\def\@ontopof#1#2#3{%
{\mathchoice
{\@@ontopof{#1}{#2}{#3}\displaystyle\scriptstyle}%
{\@@ontopof{#1}{#2}{#3}\textstyle\scriptstyle}%
{\@@ontopof{#1}{#2}{#3}\scriptstyle\scriptscriptstyle}%
{\@@ontopof{#1}{#2}{#3}\scriptscriptstyle\scriptscriptstyle}%
}%
}

\def\@@ontopof#1#2#3#4#5{%
\setbox0=\hbox{$#4#1$}%
\setbox1=\hbox{$#5#2$}%
\setbox2=\hbox{}\ht2=\ht0 \dp2=\dp0 %
\ifdim\wd0>\wd1 %
\setbox1=\hbox to\wd0{\hss\box1\hss}%
\mathord{\rlap{\raise#3\ht0\box1}\box0}%
\else   %
\setbox1=\hbox to.9\wd1{\hss\box1\hss}%
\setbox0=\hbox to\wd1{\hss$#4\relax#1$\hss}%
\mathord{\rlap{\copy0}\raise#3\ht0\box1}%
\fi
}%

\def\lambdabar{\protect\@lambdabar}
\def\@lambdabar{%
\relax
\bgroup
\def\@tempa{\hbox{\raise.73\ht0
\hbox to0pt{\kern.25\wd0\vrule width.5\wd0
height.1pt depth.1pt\hss}\box0}}%
\mathchoice{\setbox0\hbox{$\displaystyle\lambda$}\@tempa}%
{\setbox0\hbox{$\textstyle\lambda$}\@tempa}%
{\setbox0\hbox{$\scriptstyle\lambda$}\@tempa}%
{\setbox0\hbox{$\scriptscriptstyle\lambda$}\@tempa}%
\egroup
}

\def\corresponds{{\lower.2ex\hbox{=}}{\rm\kern-.75em^\triangle}}
\def\succsim{\succ\kern-.9em_\sim\kern.3em}
\def\precsim{\prec\kern-1em_\sim\kern.3em}
\def\slantfrac#1#2{\kern1em^{#1}\kern-.3em/\kern-.1em_{#2}}

\begin{document}

\begin{center}
{\Large\bf The Helical Wiggler}
\\

\medskip

Kirk T.~McDonald
\\
{\sl Joseph Henry Laboratories, Princeton University, Princeton, NJ 08544}
\\
Heinrich Mitter
\\
{\sl Institut f\"ur Theoretische Physik, Karl-Franzens-Universit\"at Graz, 
A-8010 Graz, Austria}
\\
(Oct.~12, 1986)
\end{center}

\section{Problem}

A variant on the electro- or magnetostatic boundary value problem arises
in accelerator physics, where a specified field, say ${\bf B}(0,0,z)$, is
desired along the $z$ axis.  In general there exist static fields 
${\bf B}(x,y,z)$ that reduce to the desired field on the axis, but the
``boundary condition'' ${\bf B}(0,0,z)$ is not sufficient to insure a
unique solution.

For example, find a field ${\bf B}(x,y,z)$ that reduces to
\begin{equation}
{\bf B}(0,0,z) = B_0 \cos kz \hat{\bf x} + B_0 \sin kz \hat{\bf y}
\label{p12.1}
\end{equation}
on the $z$ axis.  In this, the magnetic field rotates around the $z$ axis
as $z$ advances. 

The use of rectangular or cylindrical coordinates leads ``naturally'' to
different forms for {\bf B}.
 One 3-dimensional field extension of (\ref{p12.1}) is the
so-called helical wiggler \cite{Blewett}, which obeys the auxiliary requirement 
that the field at $z + \delta$ be the same
as the field at $z$, but rotated by angle $k\delta$. 

\section{Solution}

\subsection{Solution in Rectangular Coordinates}

We first seek a solution in rectangular coordinates, and expect that
separation of variables will apply.  Thus, we consider the form
\begin{eqnarray}
B_x & = & f(x) g(y) \cos kz,
\\
B_x & = & F(x) G(y) \sin kz,
\\
B_z & = & A(x) B(y) C(z). 
\label{s12.3}
\end{eqnarray}

Then
\begin{equation}
\nabla \cdot {\bf B} = 0 = f' g \cos kx + F G' \sin kx + A B C',
\label{s12.4}
\end{equation}
where the $'$ indicates differentiation of a function with respect to its
argument.  Equation (\ref{s12.4}) can be integrated to give
\begin{equation}
ABC = - {f' g \over k} \sin kz + {F G' \over k} \cos kx.
\label{s12.5}
\end{equation}

The $z$ component of $\nabla \times {\bf B} = 0$ tells us that
\begin{equation}
{\partial B_x \over \partial y} = f g' \cos kz = {\partial B_y \over \partial x}
= F' G \sin kz,
\label{s12.6}
\end{equation}
which implies that $g$ and $F$ are constant, say 1.  Likewise,
\begin{equation}
{\partial B_x \over \partial z} = - f k \sin kz 
= {\partial B_z \over \partial x} = A' B C = - {f'' \over k} \sin kz,
\label{s12.7}
\end{equation}
using (\ref{s12.5}-\ref{s12.6}).
Thus, $f^{''} - k^2 f = 0$, so
\begin{equation}
f = f_1 e^{kx} + f_2 e^{-kx}.
\label{s12.8}
\end{equation}
Finally, 
\begin{equation}
{\partial B_y \over \partial z} = G  k \cos kz 
= {\partial B_z \over \partial y} = A B' C = {G'' \over k} \sin kz,
\label{s12.9}
\end{equation}
so
\begin{equation}
G = G_1 e^{ky} + G_2 e^{-ky}.
\label{s12.10}
\end{equation}

The ``boundary conditions'' $f(0) = B_0 = G(0)$ are satisfied by
\begin{equation}
f = B_0 \cosh kx, \qquad G = B_0 \cosh ky,
\label{s12.11}
\end{equation}
which together with (\ref{s12.5}) leads to the solution
\begin{eqnarray}
B_x & = & B_0 \cosh kx \cos kz, 
\\
B_y & = & B_0 \cosh ky \sin kz, 
\\
B_z & = & - B_0 \sinh kx \sin kz + B_0 \sinh ky \cos kz, 
\label{s12.14}
\end{eqnarray}
This satisfies the last ``boundary condition'' that $B_z(0,0,z) = 0$.

However, this solution does not have helical symmetry.

\subsection{Solution in Cylindrical Coordinates}

Suppose instead, we look for a solution in cylindrical coordinates 
$(r,\theta,z)$.  We again expect separation of variables, but we seek to
enforce the helical symmetry that the field at $z + \delta$ be the same
as the field at $z$, but rotated by angle $k\delta$.  This symmetry
implies that the argument $kz$ should be replaced by $kz - \theta$, and
that the field has no other $\theta$ dependence.

We begin constructing our solution with the hypothesis that
\begin{eqnarray}
B_r & = & F(r)  \cos(kz - \theta), 
\\
B_\theta & = & G(r)  \sin(kz - \theta).
\label{s12.16}
\end{eqnarray}
To satisfy the condition (\ref{p12.1}) on the $z$ axis, we first transform
this to rectangular components,
\begin{eqnarray}
B_z & = & F(r) \cos(kz - \theta) \cos\theta 
               + G(r) \sin(kz - \theta) \sin\theta,
\\
B_y & = &  - F(r) \cos(kz - \theta) \sin\theta 
               + G(r) \sin(kz - \theta) \cos\theta,
\label{s12.18}
\end{eqnarray}
from which we learn that the ``boundary conditions'' on $F$ and $G$ are
\begin{equation}
F(0) = G(0) = B_0.
\label{s12.19}
\end{equation}

A suitable form for $B_z$ can be obtained from
$(\nabla \times {\bf B})_r = 0$:
\begin{equation} 
{1 \over r} {\partial B_z \over \partial \theta} =
{\partial B_\theta \over \partial z} 
= k G \cos(kz - \theta),
\label{s12.20}
\end{equation}
so
\begin{equation}
B_z = - k r G \sin(kz - \theta),
\label{s12.21}
\end{equation}
which vanishes on the $z$ axis as desired.

From either $(\nabla \times {\bf B})_\theta = 0$ or
$(\nabla \times {\bf B})_z = 0$ we find that
\begin{equation}
F = {d(rG) \over dr}.
\label{s12.22}
\end{equation}
Then, $\nabla \cdot {\bf B} = 0$ leads to
\begin{equation}
(kr)^2 {d^2 (krG) \over d(kr)^2} + kr {d (krG) \over d(kr)}
- [1 + (kr)^2](krG) = 0.
\label{s12.23}
\end{equation}
This is the differential equation for the modified Bessel function of
order 1 \cite{Abramowitz}.  Hence,
\begin{eqnarray}
G = C {I_1(kr) \over kr} = {C \over 2} \left[ 1 + { (kr)^2 \over 8} + \cdots 
\right],
\\
F = C {dI_1 \over d(kr)} = C \left( I_0 - {I_1 \over kr}  \right) =
{C \over 2} \left[ 1 + {3 (kr)^2 \over 8} + \cdots 
\right]. 
\label{s12.26}
\end{eqnarray}

The ``boundary conditions'' (\ref{s12.19}) require that $C = 2 B_0$, so
our second solution is
\begin{eqnarray}
B_r & = & 2 B_0 \left( I_0(kr) - {I_1(kr) \over kr}  \right) \cos(kz - \theta), 
\\
B_\theta & = & 2 B_0 {I_1 \over kr} \sin(kz - \theta), 
\\
B_z & = & - 2 B_0 I_1 \sin(kz - \theta),
\label{s12.29}
\end{eqnarray}
which is the form discussed in \cite{Blewett}.

\end{document}